\documentclass{nature}
\usepackage{hyperref}
\usepackage[pdftex]{graphicx,color}
\usepackage{amsmath, amssymb}
\usepackage{graphicx}
\usepackage[english]{babel}
\usepackage{subcaption}
\usepackage{enumerate}
\usepackage{tikz}
\usetikzlibrary{positioning}

\definecolor{vale}{rgb}{0,0.5, 1.}

\newcommand{\apj}{Astrophys.~J.}

\let\oldbibliography\thebibliography
\renewcommand{\thebibliography}[1]{\oldbibliography{#1}
\setlength{\itemsep}{-5pt}}

\title{\large Reply to ``\textit{Presence of a fundamental acceleration scale in galaxies}'' and ``\textit{A common Milgromian acceleration scale in nature}''}

\author{Davi C. Rodrigues$^{1,2}$, Valerio Marra$^{1,2}$, Antonino Del Popolo$^{3,4,5}$ \&  Zahra Davari$^{6}$}

\makeatletter
\let\saved@includegraphics\includegraphics
\AtBeginDocument{\let\includegraphics\saved@includegraphics}
\renewenvironment*{figure}{\@float{figure}}{\end@float}
\makeatother

\begin{document}

\maketitle

\begin{affiliations}
 \item {\footnotesize Center for Astrophysics and Cosmology, CCE,  Federal University of Esp\'irito Santo, 29075-910, Vit\'oria, ES, Brazil.}
 \item {\footnotesize  Department of Physics, CCE,  Federal University of Esp\'irito Santo, 29075-910, Vit\'oria, ES, Brazil.}
 \item {\footnotesize  Dipartamento di Fisica e Astronomia, Universit\`a di Catania, Viale Andrea Doria 6, 95125 Catania, Italy.}
 \item {\footnotesize Institute of Modern Physics, Chinese Academy of Sciences, POB 31, Lanzhou 730000, People's Republic of China.}
 \item {\footnotesize  INFN sezione di Catania, Via S. Sofia 64, 95123 Catania, Italy.}
 \item {\footnotesize  Department of Physics, Bu Ali Sina University, Hamedan, Iran.}
\end{affiliations}

\bigskip

The correspondences by McGaugh \textit{et al}\cite{McGaugh:2018aa} and Kroupa \textit{et al}\cite{Kroupa:2018aa} question our results\cite{Rodrigues:2018duc} (hereafter R18). In essence, they state that our results are in conflict with Li \textit{et al}\cite{Li:2018tdo} (hereafter L18) and criticize the priors that we used in our analysis. We show that L18 has no implication for our results and that our priors are adequate for  our analysis.

\begin{figure} 
\centering
\includegraphics[height=8cm]{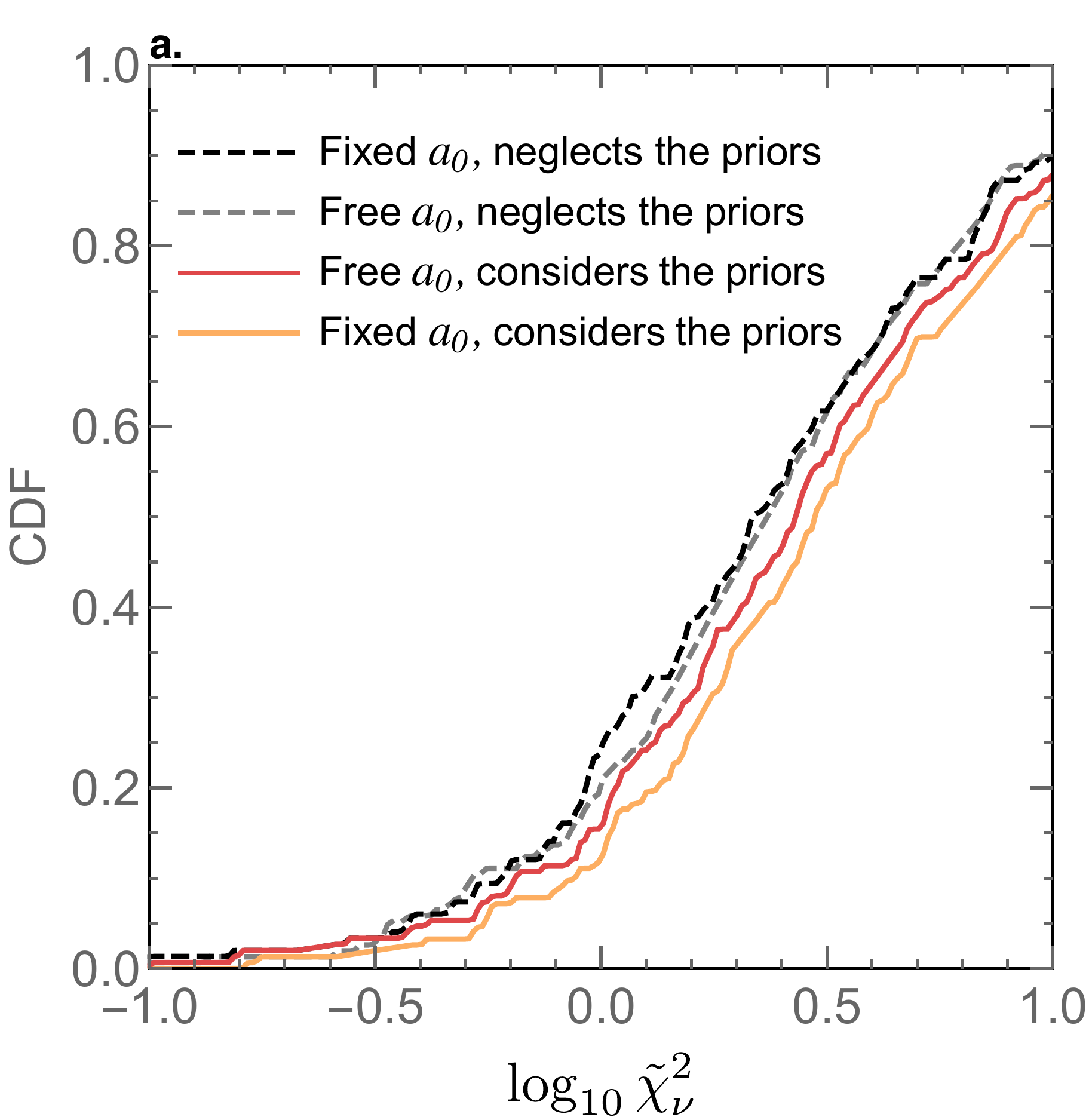} 
\includegraphics[height=8cm, trim ={0cm 0 0 0cm}, clip]{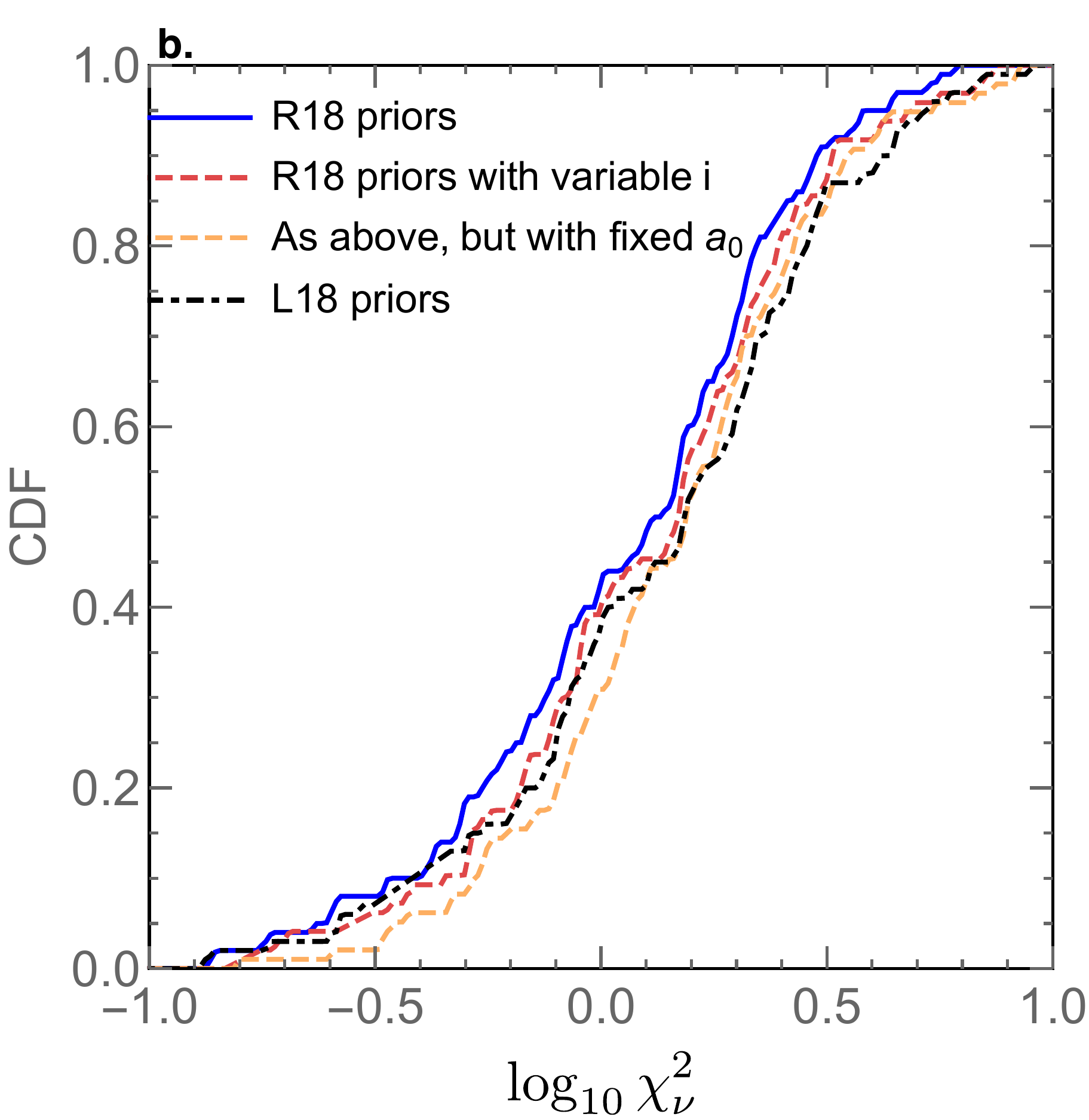}
\caption{ \small
{\bf Analyses of the $\chi^2_\nu$ CDF.} 
{\bf a.} The dashed curves reproduce the results shown in L18.
The $\tilde \chi^2$ is not the usual one, it is based on the accelerations, instead of the circular velocities\cite{Li:2018tdo}.
L18 evaluates the $\tilde \chi^2_\nu$ at the parameter values that maximize the full posterior $P$ (with priors) but does not minimize $\tilde \chi^2$. Consequently, $\tilde \chi^2_\nu$ is an effective quantity which includes no information on the priors (although they were used in the minimization process).
The same plot shows the L18 results when using another effective quantity: the minimum of $- 2 \ln P$ divided by the number of degrees of freedom (solid lines). For the latter approach, the CDF of the free $a_0$ case has an advantage.
{\bf b.}  The 100 galaxies that constitute the main analysis of R18 are  considered (we recall that galaxies with poor MONDian fits were rejected\cite{Rodrigues:2018duc}, thus the CDF achieves its maximum  faster). The standard $\chi^2$ function (the one with respect to velocities) is used here, with: $i)$ the same priors of R18, $ii)$  the priors of R18 together with inclination variations with the same Gaussian prior of L18, $iii)$ as the previous case, but with fixed $a_0 = 1.2 \times 10^{-13} \mbox{km/s}^2$, and $iv)$ all the priors of L18 with free $a_0$.  This plot follows the L18 convention when computing $\chi^2_\nu$: the priors enter only in the maximization of the posterior, not in $\chi^2_\nu$. R18 priors are favoured, but this CDF analysis is qualitative and not robust. No criticism on our results can be done based on this analysis.
}
\label{fig:chi2}
\end{figure}

L18 shows that the RAR\cite{McGaugh:2016leg} can be inferred from individual fits of galaxies, and they find no evidence for a variable acceleration scale $a_0$.
This does not contradict the fact that R18 does find strong evidence against a fundamental $a_0$. The test performed by R18 is more robust and sensitive to this  question.
Indeed, the test by L18 is based on a visual comparison of the cumulative distribution functions (CDF) of $\chi^2_\nu$ (reduced $\chi^2$).
The idea is to see how the overall fitting performance is improved when $a_0$ becomes a free parameter.
However, this test is not suitable to infer the compatibility among the $a_0$ values from different galaxies, and it is not correct to classify this approach as a Bayesian statistical analysis, as done in the McGaugh \textit{et al} correspondence.
The definitions of $\chi^2$ and $\chi^2_\nu$ used in L18 are not standard and doing a simple variation on their definition of $\chi^2_\nu$ suggests a different conclusion. Interestingly, although we do not support this CDF test, their approach applied to the main sample used in R18 favours our priors (see Figure \ref{fig:chi2}). No criticism on our results can be done from this CDF analysis.

\begin{figure}
\centering
\includegraphics[height=7cm]{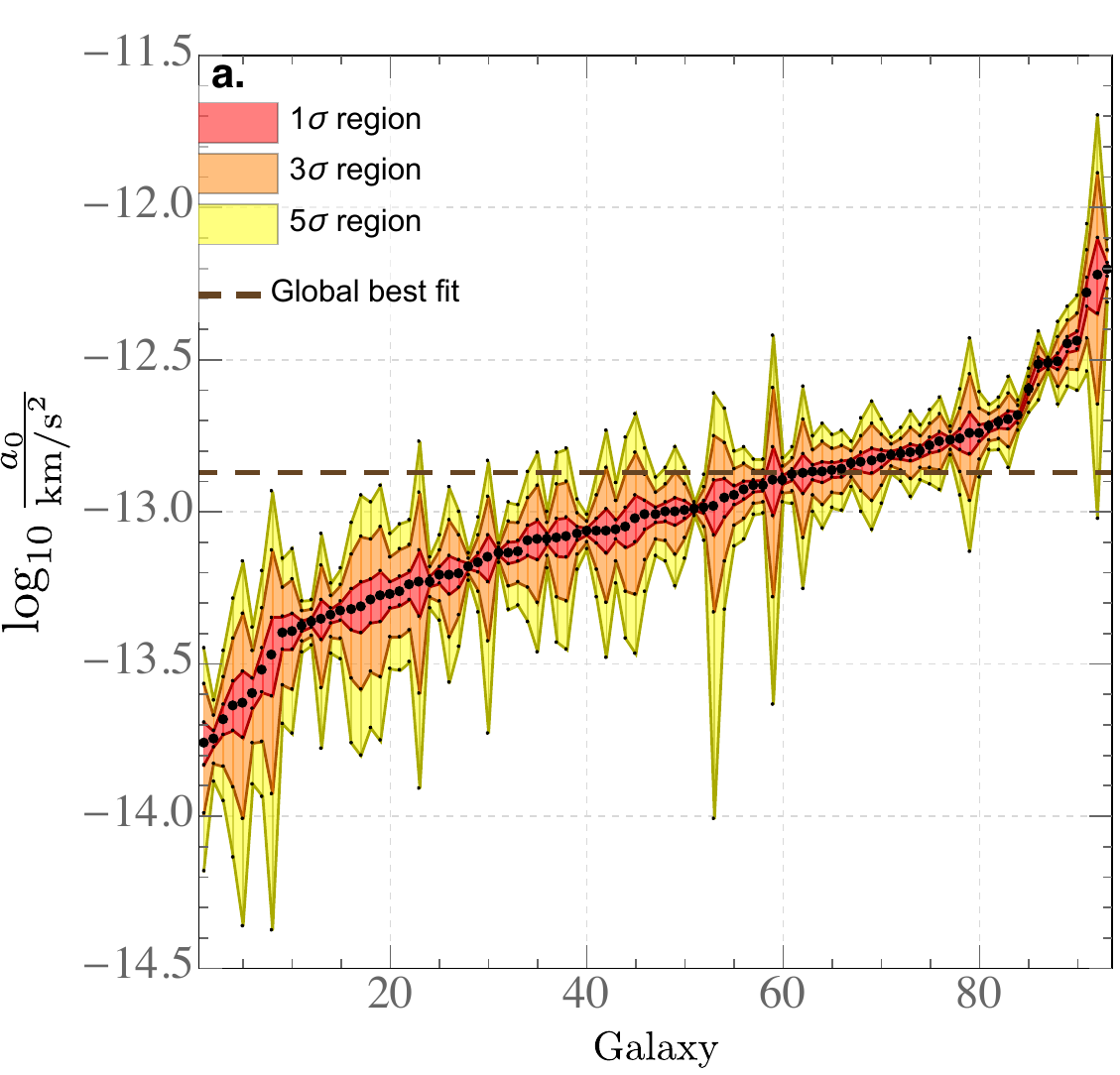} \hspace*{-0.15cm}
\includegraphics[height=7cm, trim ={2.35cm 0 0 0cm}, clip]{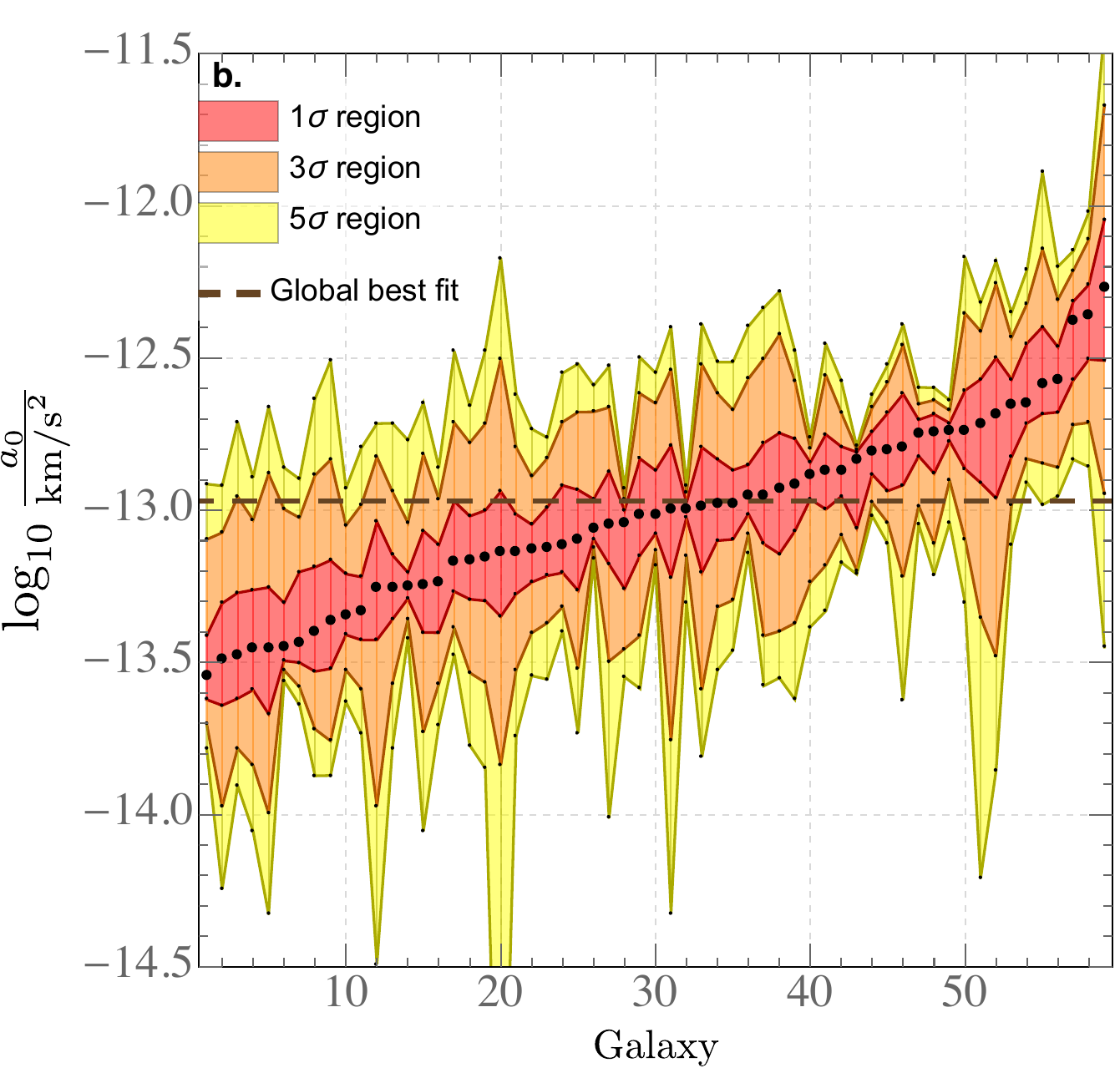} \\
\vspace*{-0.10in}
\includegraphics[height=7cm]{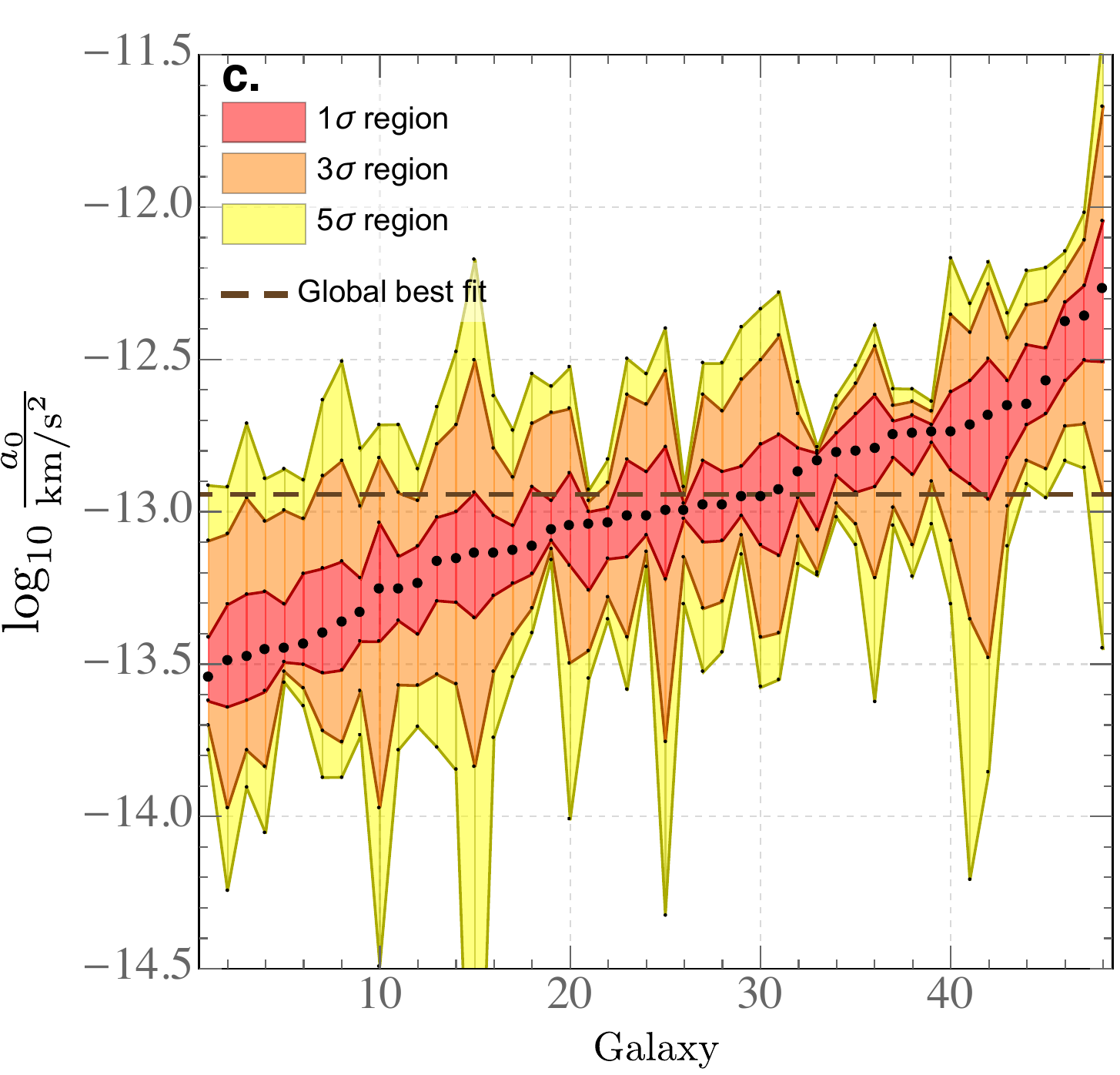} \hspace*{-0.15cm}
\includegraphics[height=7cm, trim ={3.1cm 0 0 0cm}, clip]{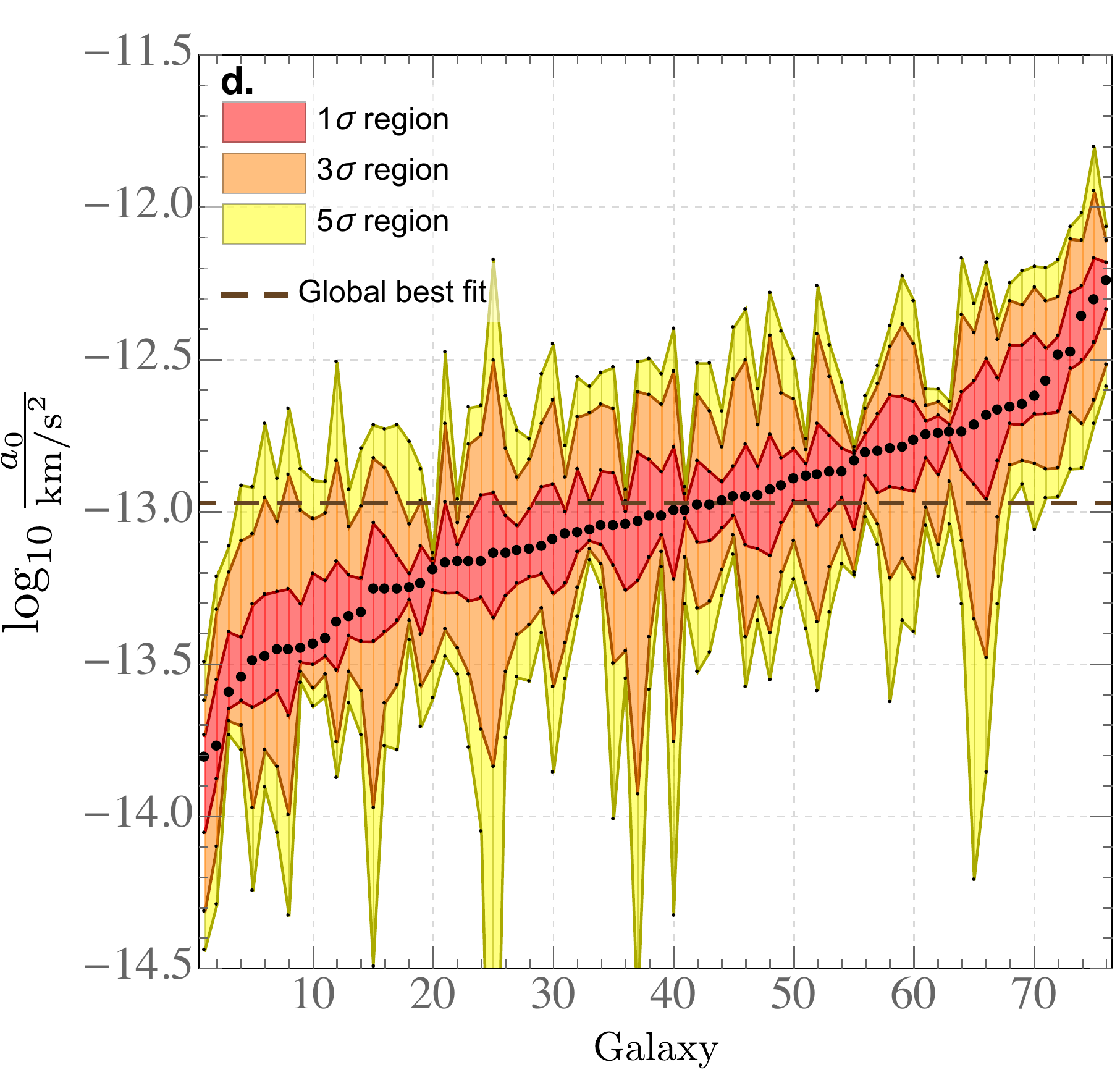}
\caption{\small
{\bf Posterior probability distributions of $a_{0}$. } The black dots are the modes of the $a_0$ posteriors (maximum probability after  marginalizing on the other parameters), the brown dashed line is the global best fit for $a_0$, the red, orange and yellow regions show the 1, 3, and $5 \sigma$ credible intervals\cite{Rodrigues:2018duc}.  
{\bf a.} Similar to Figure 1 of R18, but without the nuisance parameters: 93 galaxies pass the main quality cuts, a fundamental acceleration is rejected at very high confidence, and the modes of the $a_0$ posteriors cover a similar interval in spite of the change on the priors. 
{\bf b.} As in Figure 1 of R18, but with an additional quality cut: galaxies with relative error on $D$ larger than 20\% are excluded. This leaves 59 galaxies and a fundamental acceleration is rejected at 26$\sigma$.
{\bf c.} Similar to the previous case, but excluding all the galaxies with distance inferred from the Hubble flow. This leaves 48 galaxies and a fundamental acceleration is rejected at 22$\sigma$. 
{\bf d.} R18 always excludes galaxies with inclination ($i$) less than $30^\circ$, but one can consider a stronger quality cut, such that  galaxies with $i - 2 \sigma_i  \leq 40^\circ$ are excluded, where $\sigma_i$ is the SPARC observational error. This leaves 76 galaxies and a fundamental acceleration is rejected at 41$\sigma$.}
\label{fig:cuts}
\end{figure}

In R18 we use Bayesian inference to test the compatibility between the acceleration scales $a_0$ inferred from individual galaxies. To this end, the unknown acceleration scale $a_0$ must enter as a free parameter with a flat prior and the constraint $a_0>0$.
We find unjustified the choice of a Gaussian prior on $a_0$, as advocated in part of the analysis of L18: indeed, neither the RAR or MOND predict a fundamental $a_0$ value nor its dispersion. Consequently, here we will not discuss this case further.
Stellar mass-to-light ratios ($\Upsilon_\star$) and galaxy distances ($D$) enter in our analysis as nuisance parameters, which may capture unaccounted for systematics. Whenever there is doubt about the probability distribution of a  parameter, flat priors are better than Gaussian priors as they are the most conservative option\cite{gregory2010bayesian}. In our analysis, the effect of  nuisance parameters is to enlarge the credible intervals of $a_0$. Indeed, as we show in Figure~\ref{fig:cuts}, by fixing all the nuisance parameters at their reference values\cite{McGaugh:2016leg} and repeating our analysis with free $a_0$, the dispersion of the modes of the posteriors of $a_0$ is essentially the same, but the credible intervals are much tighter.

We were  generous with the $\Upsilon_\star$ constraints: flat priors in a 3$\sigma$ interval\cite{Meidt:2014mqa} centered on the disk and bulge values that most favour the fundamental $a_0$ interpretation\cite{McGaugh:2016leg}. 
For $D$ we used  a  $20\%$ flat prior on the SPARC reference values for all the galaxies, without entering on the merits of each type of distance estimation. The  chosen tolerance  is larger than the median value of the relative uncertainties (14$\%$). Considering Kroupa \textit{et al} comments, we add that flat priors provide larger credible intervals than using the corresponding Gaussian priors.
Also, if a galaxy cannot be well fitted with this constraint, it is  removed from our sample due to our quality cuts\cite{Rodrigues:2018duc}. That said,  it is worthwhile to evaluate the dependence of our results on the galaxies with the largest relative uncertainties, this is shown in  Figure \ref{fig:cuts}. By removing either galaxies with distance uncertainties larger than 20$\%$ or those measured from the Hubble flow, a fundamental acceleration scale is still rejected at more than 10$\sigma$ (see also R18 for further details).

One can use an unlimited number of nuisance parameters to describe a galaxy, we selected the most relevant ones and found a result with high level of statistical significance (for all our quality cuts, it was always beyond 10$\sigma$, and for our main analysis, we find $\sim 50 \sigma$). No reasonable inclination variation can wash away those 50$\sigma$, since by eliminating all the galaxies more susceptible to inclination changes our conclusion is the same, see Figure \ref{fig:cuts}. At last, we also recall that we analysed 18 galaxies from the THINGS sample, and we found similar results.

\noindent
{\bf On the histograms presented in the McGaugh \textit{et al} correspondence.} Our results\cite{Rodrigues:2018duc} are  not based on best fits. A  histogram on these quantities can argue in favour of the emergent nature of the RAR but it does not directly test the compatibility between the acceleration scales.
We also add that the analysis of McGaugh \textit{et al} does not consider the same sample of data that we used to infer our conclusions, since our quality cuts were not considered. We use their suggestion of eliminating galaxies with large distance uncertainties in Figure \ref{fig:cuts}.  

\noindent
{\bf On the quality cuts of Kroupa \textit{et al} and inclinations fits.} After a sequence of cuts on the sample from which the RAR\cite{McGaugh:2016leg} is derived, 81\% of the RAR sample is removed and still 17\% of the remaining galaxies are incompatible with MOND, according to their criteria.  {\it A posteriori} adjustments on particular features of each galaxy are evoked such that compatibility with MOND, in some sense, is achieved. We do not find that this is a good argument in favour of MOND. If there is reason to suspect of observational data issues, all the galaxies need to be reevaluated, not only those with problems with MOND. Among their cuts, the strongest one is the Hubble flow one, which we used in Figure \ref{fig:cuts}. At last, the plot with best fits on inclinations does not address the issue of compatibility between acceleration scales, nor any probability has been quantified.

\begin{addendum}

\item DCR and VM thank CNPq and FAPES (Brazil) for partial financial support. ADP was supported by the Chinese Academy of Sciences and by the President's international fellowship initiative, grant no. 2017 VMA0044. ZD thanks the ministry of science, research and technology of Iran for  financial support.

\end{addendum}



\end{document}